\documentclass[aps,pre,twocolumn,superscriptaddress,amsmath,showpacs]{revtex4}

\usepackage{graphicx, enumitem, multirow}
\usepackage[outercaption]{sidecap}

\begin{document}

%\begin{htmlonly}
%  \def\href#1#2{\htmladdnormallink{#2}{#1}}  %for L2H only
%\end{htmlonly}

\title{Impact of sequential disorder on the scaling behavior of airplane boarding time}

\author{Yongjoo Baek}
\affiliation {Department of Physics,
Korea Advanced Institute of Science and Technology, Daejeon
305-701, Korea}

\author{Meesoon Ha}
\email[Corresponding author: ]{msha@chosun.ac.kr}
\affiliation{Department of Physics Education, Chosun University,
Gwangju 501-759, Korea}

\author{Hawoong Jeong}
\email[]{hjeong@kaist.edu}
\affiliation{Department of Physics and Institute for the
BioCentury, Korea Advanced Institute of Science and Technology,
Daejeon 305-701, Korea}
\affiliation{APCTP, Pohang, Gyeongbuk
790-784, Korea}

\date{\today}

\begin{abstract}
Airplane boarding process is an example where disorder properties
of the system are relevant to the emergence of universality
classes. Based on a simple model, we present a systematic analysis
of finite-size effects in boarding time, and propose a
comprehensive view of the role of sequential disorder in the
scaling behavior of boarding time against the plane size. Using
numerical simulations and mathematical arguments, we find how the
scaling behavior depends on the number of seat columns and the
range of sequential disorder. Our results show that new scaling
exponents can arise as disorder is localized to varying extents.
\end{abstract}


%----------------------------------------------------------------------------
%89.75.Da Systems obeying scaling laws
%89.40.Dd Air transportation
%02.50.-r Probability theory, stochastic processes, and statistics
%05.40.-a Fluctuation phenomena, random processes, noise, and Brownian motion
%----------------------------------------------------------------------------
\maketitle

\section{Introduction}
How disorder properties affect the scaling behavior of a characteristic
time scale is a problem studied in various areas of physics. For example,
dynamic exponents of surface growth models~\cite{Barabasi1995} depend
on the geometry of substrate and the type of disorder.
In complex networks, consensus time of opinion dynamics~\cite{Sood2005}
and first-passage time of random walk~\cite{SHwang2012} scale
differently with the system size depending on the level of structural
heterogeneity. Those studies reveal the subtle interplay between
dynamics and disorder that gives rise to different universality classes
of scaling behaviors. Expanding the list of universality classes and
clarifying their origins has been established as a general framework frequently
employed by physicists.

Airplane boarding process provides another interesting example
where this framework can be applied. The average time required for
all passengers to get seated, or {\em average boarding time}
$\langle T \rangle$, may scale differently with the plane size $N$
depending on sequential disorder of passengers, which is
controlled by the airline's boarding policy and each individual's
gate arrival time. However, most studies of boarding process were
limited to the practical problem of reducing boarding time at a
fixed plane size~\cite{Marelli1998, Landeghem2002, Briel2005,
Ferrari2005, Bazargan2007, Steffen2008, Nyquist2008}, which
provides only fragmented knowledge about boarding time in
prescribed situations. To understand the general nature of
boarding process, we should examine its scaling properties, which
came to be studied only very recently. Analytical studies by
Bachmat {\em et al.}~\cite{Bachmat2006, Bachmat2007, Bachmat2008,
Bachmat2009} argued for $\langle T \rangle \sim N^{1/2}$ on the
basis of mathematical results about longest monotonic subsequences
in random permutations of real number pairs~\cite{Vershik1977,
Deushel1995}. Meanwhile, a numerical study by Frette and
Hemmer~\cite{Frette2012} based on a simple boarding model (see
Fig.~\ref{fig:boarding}) reported $\langle T \rangle \sim
N^{0.69}$. The conflict between the two exponents has been
discussed by subsequent numerical works, which supported $\langle
T \rangle \sim N^{1/2}$ as the true asymptotic behavior after the
finite-size effect is systematically filtered
out~\cite{Bernstein2012,Brics2013}.

In this paper, we revisit the original mathematical
theorem~\cite{Vershik1977, Deushel1995} that led to the prediction
of $N^{1/2}$ scaling. A physical interpretation of its conditions
reveals that previous
studies~\cite{Bachmat2006,Bachmat2007,Bachmat2008,Bachmat2009,
Frette2012,Bernstein2012,Brics2013} were limited to the cases with
a broad range of disorder in the passenger sequence, and that
different scaling behaviors are possible if the range of
sequential disorder is localized to various extents. This leads us
to a more comprehensive picture of the scaling behaviors of
$\langle T \rangle$.

This paper is organized as follows. In Sec.~\ref{sec:model}, we
generalize the boarding model proposed by~\cite{Frette2012},
incorporating additional parameters for airplane structure and
sequential disorder. In Sec.~\ref{sec:numerical}, we make
analytical predictions on possible scaling behaviors in special
cases of fully reserved planes. Our predictions are numerically
confirmed in Sec.~\ref{sec:numerical}, where they are also verified
for more complicated situations, taking into account factors
such as vacancy effect and general localized disorder.
Finally, we conclude with a summary of our findings in
Sec.~\ref{sec:summary}.

\section{Model}
\label{sec:model}
\subsection{Boarding process}
\begin{figure}[t]
\centering
\includegraphics[width=\columnwidth]{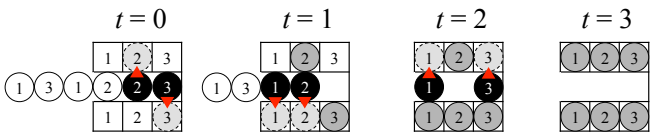}
\caption{\label{fig:boarding} (Color online) Snapshots of boarding
in an airplane with $N = 3$ rows and $c = 2$ columns of seats,
where each passenger (circle) is labeled by the row index of its
seat. At each time step, passengers are waiting in line (white),
sitting down (black), or staying seated (gray).}
\end{figure}

We consider an airplane with $N$ rows and $c$ columns of seats
along a one-dimensional aisle. Each seat is labeled with a row index $r$
if it is in the $r$-th row from the front. The aisle is discretized into $N$ sites,
each of which cannot hold more than one passenger at once.
At $t = 0$, $cN$ passengers enter the plane from the front in a sequence
$\left\{r_1, r_2, \ldots, r_{cN}\right\}$, where $r_i$ denotes the row index
of the $i$-th passenger to enter.
Passengers' positions are synchronously updated. Every passenger
moves along the aisle front-to-back, instantly crossing successive
empty sites until blocked by an occupied site or reaching the row of
its assigned seat. If the latter is the case, sitting down takes one time
step for every passenger. Continuing the process, all passengers would
get seated at $t = T$, which we call boarding time (see Fig.~\ref{fig:boarding}).

\begin{figure}[b]
\centering
\includegraphics[width=\columnwidth]{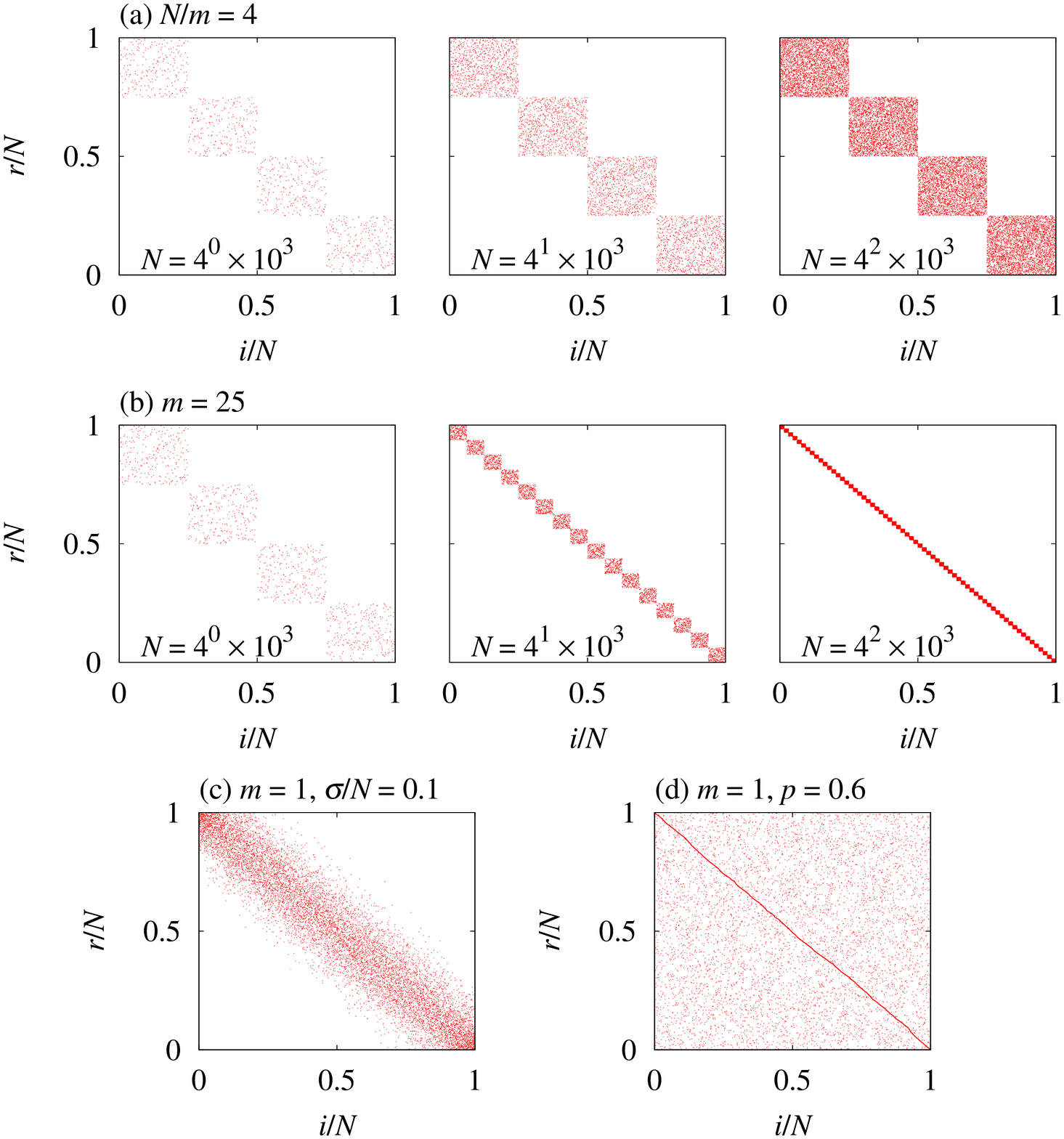}
\caption{\label{fig:shuffling} (Color online) Examples of the
passenger sequence $\{r_i\}$. As $N \to \infty$, (a) fixed $N/m$
means globalized disorder; (b) fixed $m$ means localized disorder;
(c) Gaussian shuffling gives either purely globalized or purely
localized disorder; (d) uniform shuffling allows coexistence of
both kinds of disorder. We set $N = 10^4$ in (c) and (d).}
\end{figure}

\subsection{Sequential disorder}
Although $T$ is completely determined by the initial passenger sequence,
the entry order of each passenger is not strictly controlled in normal situations.
Thus, it makes more sense to deal with $\langle T \rangle$ for an ensemble
of initial passenger sequences.

The ensemble of sequences is primarily constrained by boarding policies.
Here we focus on back-to-front policies with equal-sized boarding groups.
According to such policies, passengers are divided into boarding groups
of $m$ rows each ($m$ is a divisor of $N$),
so that the $n$-th group to enter the plane consists of passengers
whose row indices satisfy $N-nm+1 \le r_i \le N-(n-1)m$.
Consequently, the passenger sequence is sorted back-to-front on a
scale larger than groups, but remains randomized within each group.
Thus, $m$ can be interpreted as the range of sequential disorder.
If $N/m$ ($m$) is fixed as $N \to \infty$, we can say that
sequential disorder is purely globalized (localized),
as illustrated in Fig.~\ref{fig:shuffling}(a) and (b), respectively.

Meanwhile, some passengers would deviate from the boarding policy
due to their unpunctuality. This justifies the incorporation of
arrival time fluctuations as another determinant of sequential
disorder. We consider Gaussian shuffling as a model of such
fluctuations, which is defined as follows: we add an independent
and identically distributed Gaussian random variable $\eta_i$ of
zero mean and of variance $\sigma^2$ to the sequential index $i$
of each passenger, and then sort the sequence in the increasing
order of $i+\eta_i$ [see Fig.~\ref{fig:shuffling}(c)]. As long as
$m$ is finite, Gaussian shuffling makes sequential disorder purely
localized (globalized) if $\sigma$ ($\sigma/N$) is fixed in the
asymptotic limit. Alternatively, we can also consider uniform
shuffling, in which every passenger is randomly relocated in the
sequence with probability $p$ [see Fig.~\ref{fig:shuffling}(d)].
If $m$ is finite, uniform shuffling allows localized and
globalized disorder to coexist until the sequence is completely
randomized at $p = 1$.

\section{Theoretical predictions}
\label{sec:theory}
\subsection{Scaling behaviors for globalized disorder}
As illustrated in Fig.~\ref{fig:shuffling}, a passenger sequence
can be rendered as a two-dimensional scatter plot, horizontal
and vertical axes representing the sequential index $i$ and the
row index $r$, respectively. In the asymptotic limit $N \to
\infty$, an ensemble of sequences becomes equivalent to a
probability density function (PDF) $p(i/N,r/N)$ defined over a
continuous two-dimensional area~\cite{Bachmat2006, Bachmat2007,
Bachmat2008, Bachmat2009}. This representation enables us to
utilize the following mathematical theorem~\cite{Vershik1977,
Deushel1995}.

{\em Theorem} --- If $(x_\alpha, y_\alpha)$, $\alpha = 1, \ldots,
N$ are pairs of real numbers with $0 \le x_\alpha \le 1$ and $0
\le y_\alpha \le 1$, we say that a subsequence $\left\{(x_{i_1}, y_{i_1}),
\ldots, (x_{i_l}, y_{i_l})\right\}$ is an {\em increasing subsequence} if
\begin{equation*}
x_{i_j} < x_{i_{j+1}} \text{ and } y_{i_j} < y_{i_{j+1}} \text{ for } j = 1,\ldots,l-1
\end{equation*}
where $i_j$ is a sequence of non-repeated indices between $1$
and $N$. If the pairs $(x_\alpha, y_\alpha)$ are generated from a
finite PDF $p(x,y)$, the length of the longest increasing subsequence
asymptotically scales as $N^{1/2}$.

The increasing subsequence in the theorem can be translated as
the blocking subsequence of passengers, in which one passenger
{\em blocks} the next if the latter cannot reach its seat unless
the former is seated~\cite{Bachmat2006, Bachmat2009}. Since the
length of the longest blocking subsequence is equal to boarding
time $T$, the theorem indicates that $T$ (and $\langle T \rangle$
as well) scales as $N^{1/2}$ if the PDF remains finite in the asymptotic
limit $N \to \infty$. The finitude of the PDF is equivalent to the broad
range of disorder in the passenger sequence, which corresponds to
globalized disorder defined in the previous section.
In other words, $\langle T \rangle \sim N^{1/2}$ is guaranteed for
purely globalized disorder.

\begin{figure*}[]
\centering
\includegraphics[width=0.8\textwidth]{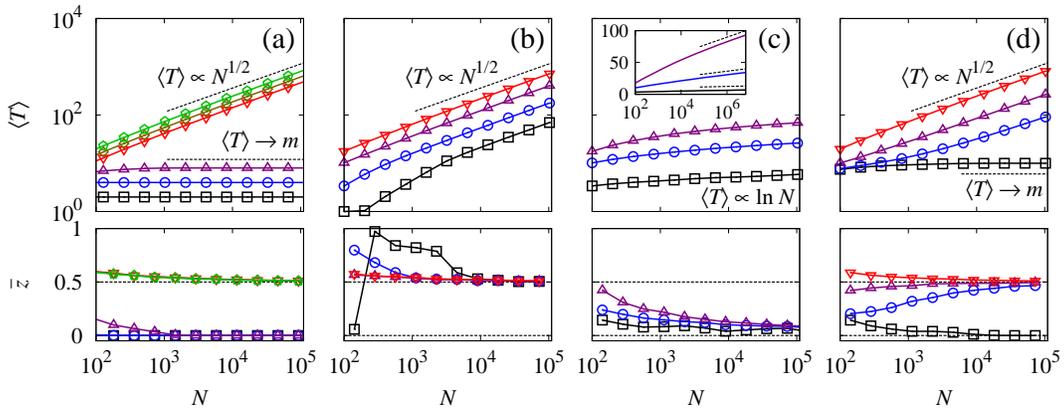}
\caption{\label{fig:single} (Color online) Asymptotic scaling
behaviors of $\langle T\rangle$ in a fully reserved plane with
$c=1$. (Upper panel) From bottom to top, the curves represent (a)
$m = 2,4,8,N/4,N/2,N$ without shuffling, and $m = 10$ with (b)
$\sigma /N = 0.001,0.01,0.1,1$, (c) $\sigma = 1,10,100$ (inset:
semi-log plots of the main), (d) $p = 0,0.01,0.1,1$. (Lower panel)
Effective scaling exponents $\bar{z}$ of corresponding curves in
the upper panel. Each data point is averaged over at least $100$
samples.}
\end{figure*}

\begin{figure*}
\centering
\includegraphics[width=0.8\textwidth]{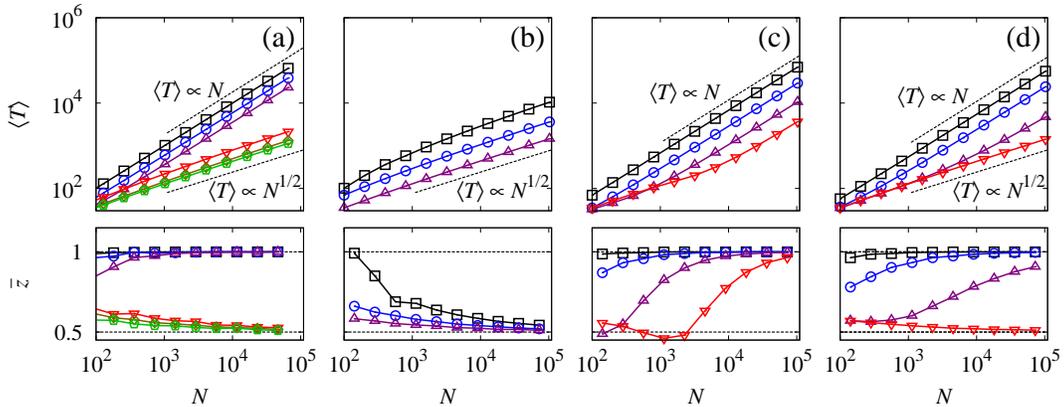}
\caption{\label{fig:multi} (Color online) Asymptotic scaling
behaviors of $\langle T\rangle$ in a fully reserved plane with
$c=2$. (Upper panel) From top to bottom, (a) $m =
1,4,16,N/16,N/4,N$ without shuffling, and $m = 5$ with (b) $\sigma
/N = 0.001,0.01,0.1$, (c) $\sigma = 1,10,100,1000$, (d) $p =
0,0.5,0.9,1$. (Lower panel) Effective scaling exponents $\bar{z}$
of corresponding curves in the upper panel. Each data point is
averaged over at least $100$ samples.}
\end{figure*}

\subsection{Scaling behaviors for localized disorder}
Meanwhile, different scaling behaviors might arise if some part of
the PDF diverges due to the presence of localized disorder
in the passenger sequence. Here we present theoretical arguments
on the scaling behaviors of $\langle T \rangle$ for fully reserved planes
when $m$ is kept finite without arrival time fluctuations.

\subsubsection{Single-column planes ($c = 1$)}
We first consider a fully reserved plane with a single column ($c
= 1$) of seats. When a boarding policy fills the window seats
first and the aisle seats later, there is effectively no
interference between the passengers to be seated in the same row.
In such a case, even a multicolumn plane can be regarded as a
single-column plane.

We start with an argument on a lower bound of $T$. As $N \to
\infty$, the passenger sequence always contains a boarding group
whose row-index configuration is exactly front-to-back, e.g.,
$\left\{1,2,\ldots,m-1,m\right\}$. Since every passenger blocks
the next passenger, boarding time for this particular group is
$m$. Thus, we have $\lim_{N\to\infty} T \ge m$.

Now we turn to an upper bound of $T$. We assume the worst-case
scenario: a group starts to board when $m-1$ seats out of $m$
seats reserved by its members are blocked by the previous group,
and the number of blocked seats decreases by one at each time
step. We choose an arbitrary passenger who has $n_{<}+n_{>}-1$
sites to go before reaching its seat, where $n_{<}$ ($n_{>}$)
denotes the number of passengers in the same group whose row
indices are smaller (larger). Those $n_{<}$ passengers would
eventually get seated before the chosen passenger, leaving empty
sites in the aisle. Whenever such empty sites appear, the chosen
passenger instantly advances along the aisle by as many sites in
addition to the base speed of one site per unit time. Hence, the
chosen passenger can reach the reserved seat within $n_{>}-1$ time
steps, spending one more time step to sit down. Since $n_{>} \le
m$, individual boarding time for any passenger is not greater than
$m$, and so is the total boarding time, i.e., $T \le m$. Combining
both upper and lower bounds, we obtain $\lim_{N\to\infty} T = m$.

\subsubsection{Multicolumn planes ($c \ge 2$)}
As a next step, we consider a fully reserved plane with multiple columns
($c \ge 2$) of seats. This is the typical situation encountered in reality,
provided that passengers belonging to different columns are allowed to
mix together while entering the plane.

We again begin with a lower bound of $T$. The number of boarding
groups whose row-index configuration is $c$ repetitions of
$\left\{1,2,\ldots,m\right\}$ grows linearly with $N$. When the
first $m$ members of such a group begin to take seats, the rest of
the group completely occupy the rows reserved for the next group.
Thus, the next group cannot reach their seats for a finite number
of time steps, which implies that $T$ increases by one or more
time steps for every occurrence of this configuration. Thus, $T
\ge aN$ as $N \to \infty$, where $a$ is a positive constant.

The total number of passengers $cN$ is trivially an upper bound of $T$,
since there is always at least one passenger sitting down per unit time.
Therefore, we have $\langle T \rangle \sim N$.

\begin{figure}[t]
\centering
\includegraphics[width=0.9\columnwidth]{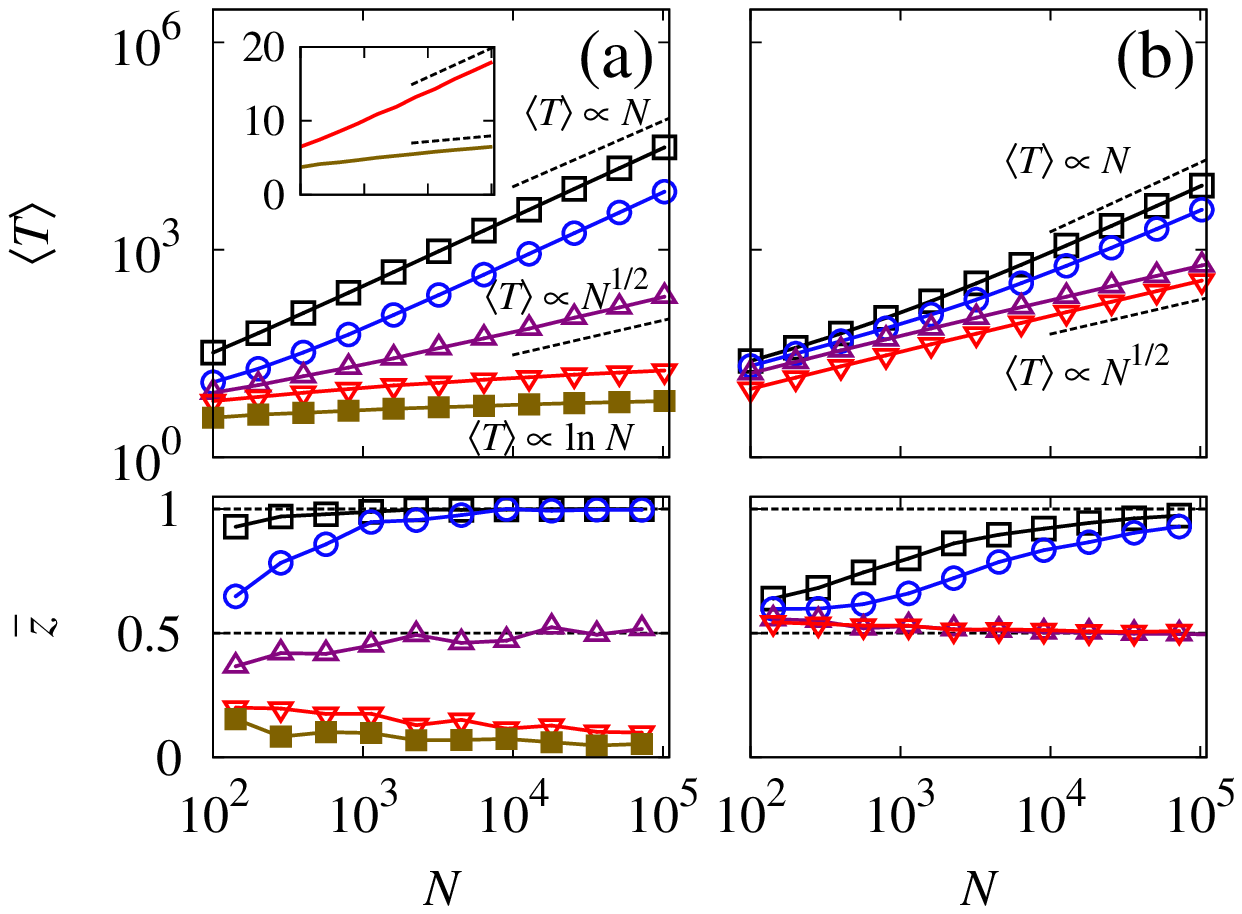}
\caption{\label{fig:scaling_phi} (Color online) Asymptotic scaling
behaviors of $\langle T \rangle$ in a partially reserved
two-column ($c=2$) airplane with group size $m = 5$. (Upper panel)
For the curves from top to bottom, (a) $\phi = 0.25, 0.45, 0.5,
0.55, 0.75$ without shuffling (inset: semilog plots of the lowest
two curves in the main) and (b) $\phi = 0.25, 0.35, 0.5, 0.55$
with uniform shuffling $p = 0.5$. (Lower panel) Effective scaling
exponents $\bar{z}$ of corresponding curves in the upper panel.
Each data point is averaged over at least $100$ samples.}
\end{figure}

\section{Numerical results}
\label{sec:numerical}
In this section, we numerically test the scaling behaviors of $\langle T \rangle$
discussed in the previous section and check their validity in more general cases.
To clarify the true asymptotic behavior of $\langle T \rangle$, we use effective
scaling exponent $\bar{z}$ defined as follows~\cite{Bernstein2012,Brics2013}:
\begin{equation*}
\bar{z}(\sqrt{N_i N_{i+1}}) \equiv \frac{\ln
\left[ \langle T\left(N_{i+1}\right) \rangle / \langle T\left(N_{i}\right)
\rangle \right]}{\ln \left(N_{i+1}/N_{i}\right)}.
\end{equation*}
This is none other than the average slope of $\langle T \rangle$
between $N_i$ and $N_{i+1}$ in the log-log plot of $\langle T
\rangle$ against $N$. We interpret the saturation value of
$\bar{z}$ in the asymptotic limit as the true scaling exponent of
$\langle T \rangle$.

\subsection{Single-column planes ($c = 1$)}
The scaling behaviors of $\langle T \rangle$ for $c = 1$ are shown
in the upper panel of Fig.~\ref{fig:single} and the behaviors of
$\bar{z}$ in the lower panel. Without arrival time fluctuations,
$\langle T \rangle$ scales as $N^{1/2}$ if $N/m$ is fixed, while
it saturates to $m$ if $m$ is fixed [see
Fig.~\ref{fig:single}(a)]. Gaussian shuffling at finite $m$
produces similar but slightly different scalings: $N^{1/2}$
scaling for fixed $\sigma/N$ [see Fig.~\ref{fig:single}(b)] and
$\ln N$ scaling for fixed $\sigma$ [see Fig.~\ref{fig:single}(c)].
Since both saturation and log scaling are slower than any
algebraic scaling, they can be collectively labeled as $N^0$
scalings. Thus, our results confirm $\langle T \rangle \sim
N^{1/2}$ for purely globalized disorder, while finding $\langle T
\rangle \sim N^0$ for purely localized disorder, regardless of the
presence of arrival time fluctuations. When those two kinds of
disorder coexist, the effect of localized disorder predominates,
as implied by $N^{1/2}$ scaling observed whenever $p > 0$ [see
Fig.~\ref{fig:single}(d)].

Before moving on to the multicolumn case, we remark that dividing
into smaller groups always reduces boarding time if $c = 1$, as
previously reported by \cite{Bernstein2012}. However, we emphasize
that the benefits of group division, in terms of scaling, are not
very robust against disorder caused by arrival time fluctuations.
A slightest hint of globalized disorder can revert the scaling
behavior to that of the random boarding policy.

\subsection{Multicolumn planes ($c \ge 2$)}
Without loss of generality, we focus on the boarding process of
a two-column ($c = 2$) plane since the scaling behaviors does not
change in the other cases.

Figure~\ref{fig:multi} shows the scaling behaviors of $\langle T
\rangle$ at $c = 2$. If no shuffling is applied, $\langle T
\rangle$ scales as $N^{1/2}$ ($N$) if $N/m$ ($m$) is fixed [see
Fig.~\ref{fig:multi}(a)]. Gaussian shuffling at finite $m$ results
in exactly the same scalings [see Fig.~\ref{fig:multi}(b) and
(c)], implying $\langle T \rangle \sim N^{1/2}$ ($\langle T
\rangle \sim N$) for purely globalized (localized) disorder, even
in the presence of arrival time fluctuations. If both kinds of
disorder are mixed together, the effect of globalized disorder
predominates, as linear scaling for $p < 1$ indicates [see
Fig.~\ref{fig:multi}(d)].

We note that division into smaller groups always increases
boarding time if $c \ge 2$. In this case, reducing the
range of disorder always results in stronger interference between
passengers to be seated near each other, as reflected in the
strong sensitivity of the scaling exponent to localized disorder.
Thus, lack of control is better than row-wise division of groups
for $c \ge 2$ planes (provided that we do not divide passengers
column-wise, which was covered by the $c = 1$ case), which is a
lesson shared by most of previous studies~\cite{Landeghem2002,
Briel2005, Ferrari2005, Bazargan2007, Steffen2008, Nyquist2008,
Bachmat2006, Bachmat2007, Bachmat2008, Bachmat2009}.

\subsection{Effect of vacancy}
We also consider the asymptotic scaling behaviors of $\langle T
\rangle$ for $c = 2$ [see Fig.~\ref{fig:scaling_phi}] when each
seat can be vacant with probability $\phi$. A similar case was
also studied by \cite{Ferrari2005}, but only at a fixed plane size
without any consideration of scaling.

Figure~\ref{fig:scaling_phi}(a) shows the effect of $\phi$ when
there are no arrival time fluctuations. $\langle T \rangle$
scales as $N$ for $\phi < 1/2$ and as $\ln N$ for $\phi > 1/2$,
implying that $\phi$ can interpolate between the scalings for
localized disorder observed at $c = 1$ and $c \ge 2$. The same
observation can be made even in the presence of globalized
disorder, as shown in Fig.~\ref{fig:scaling_phi}(b), where the
scaling changes from $\langle T \rangle \sim N$ to $\langle T
\rangle \sim N^{1/2}$. The transition point of the scaling
exponent is $1/2$ for $c = 2$, which is confirmed to be $\phi_c =
1 - 1/c$ for the general value of $c$ (even including $c = 1$, in
which case the scaling is not affected by vacancy). Interestingly,
$\langle T\rangle \sim N^{1/2}$ seems to hold exactly at $\phi =
\phi_c$, regardless of the nature of sequential disorder.

\subsection{Generalization of localized disorder}
Finally, we generalize the range of sequential disorder even further
by applying Gaussian shuffling with $\sigma \sim N^s$,
where $s$ is freely varied between $0$ and $1$.
This allows us to interpolate between purely localized and
purely globalized disorders without mixing different kinds of disorders
as in the case of uniform shuffling.

\begin{figure}[t]
\centering
\includegraphics[width=\columnwidth]{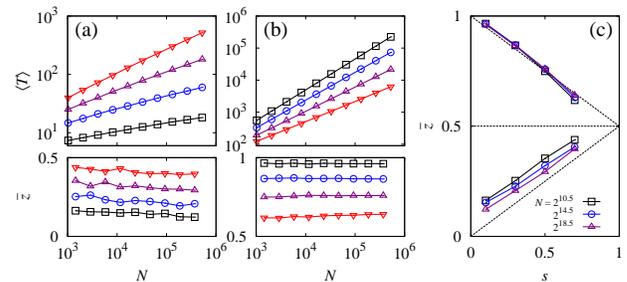}
\caption{\label{fig:general} (Color online) Asymptotic scaling
behaviors of $\langle T \rangle$ when the range of disorder is
given by Gaussian shuffling with $\sigma = N^s$. In the upper
panel, (a) from bottom to top, $c = 1$ with $s = 0.1,0.3,0.5,0.7$;
(b) from top to bottom, $c = 2$ with $s = 0.1,0.3,0.5,0.7$.
Behaviors of $\bar{z}$ for corresponding curves in the upper
panels are plotted in the lower panels of (a) and (b). (c)
$\bar{z}$ as a function of $s$ for different values of $N$. Each
data point is averaged over at least $100$ samples.}
\end{figure}

$N$ dependence of $\langle T \rangle$ and $\bar{z}$ for $c = 1$
and $c = 2$ are plotted in Fig.~\ref{fig:general}(a) and (b),
respectively. It is clear that the scaling exponent of $\langle T
\rangle$ changes continuously as $s$ is varied, monotonically
increasing (decreasing) with $s$ if $c = 1$ ($c = 2$), as
indicated by the $s$-dependence of $\bar{z}$ in
Fig.~\ref{fig:general}(c). The results hint at a possible linear
dependence of the scaling exponent on $s$ [represented by dashed
lines in Fig.~\ref{fig:general}(c)], which remains a speculation
at the moment.

\section{Summary and discussions}
\label{sec:summary} 
\begin{figure}
\includegraphics[width=\columnwidth]{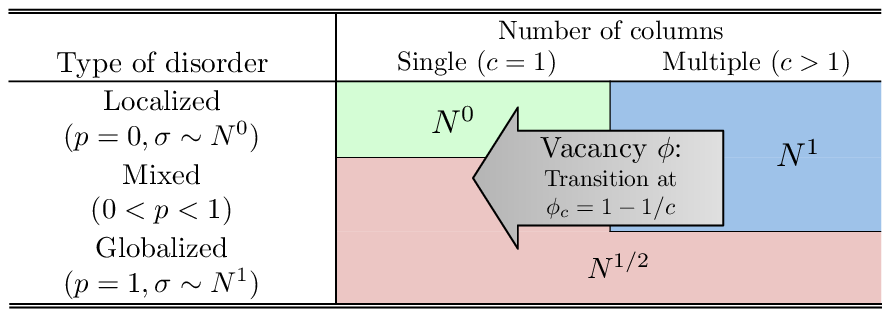}
\caption{\label{tab:scaling_table} (Color online) Summary of
asymptotic scalings of $\langle T \rangle$. Note that $N^0$ stands
for both saturation to a finite value and logarithmic scaling. The
gray arrow indicates that vacancy effect changes the multicolumn
scaling to the single-column one at $\phi_c$.}
\end{figure}

Scaling behaviors of average boarding time are
summarized in Fig.~\ref{tab:scaling_table}. In a single-column
($c=1$) plane, purely localized disorder produces $N^0$ scaling
including both saturation and logarithmic divergence, and purely
globalized disorder leads to $N^{1/2}$ scaling. The boarding
process is more sensitive to globalized disorder, so $N^{1/2}$
scaling is observed when both kinds of disorder are present. On
the other hand, in a multicolumn ($c \ge 2$) plane, purely
localized disorder produces linear scaling, while purely
globalized one again yields $N^{1/2}$ scaling. Since localized
disorder is dominant in this case, linear scaling is observed when
both kinds of disorder are present. Increasing the probability of
vacant seats $\phi$ beyond $\phi_c = 1 - 1/c$ changes scaling
behaviors from the multicolumn ones to the corresponding
single-column ones while keeping sequential disorder type.
However, the borderline scaling at $\phi_c$ seems to be $N^{1/2}$,
regardless of disorder type. Expanding
Fig.~\ref{tab:scaling_table} to encompass general localized
disorder with noninteger $s$ remains a challenge for future works.

To sum up, we have systematically investigated the relationship
between the range of disorder in the passenger sequence and the
asymptotic scaling behavior of boarding time using a simple
boarding model. Our results clarify the origins of different
scalings and indicate which type of disorder plays a dominant
role. This offers a natural way to incorporate the boarding
problem, previously regarded as an engineer's optimization
problem, into the domain of physics.

Finally, we add a few remarks on other possible generalizations of
our study. For example, we can allow fluctuations in the time
required for each passenger to sit down. Since such fluctuations
do not affect the scaling behavior of the longest blocking
subsequence, they would not affect the scaling behaviors as long
as they are finite and uncorrelated, whereas diverging or
correlated fluctuations may produce interesting changes. We also
note that the model considered in our study is similar to
asymmetric simple exclusion process (ASEP)~\cite{ASEP}. It would
be interesting to check how concepts of ASEP apply to the case of
boarding as differences like hopping rates of particles, diversity
of destinations, and synchronous update are removed to varying
extents. We hope these points to be satisfactorily addressed in
future studies.

\section*{ACKNOWLEDGMENTS}
The work was supported by the National
Research Foundation of Korea (NRF) grant funded by the Korean
Government (MEST) (No. 2012-0004216) (M.H.); (No. 2011-0028908)
(Y.B., H.J.). M.H. would also acknowledge the generous hospitality
of KIAS for Associate Member Program, funded by the MEST.

\end{document}